\documentclass[showpacs,twocolumn,prd,nofootinbib]{revtex4-1}
 \usepackage{color,graphicx,epsfig}
 \usepackage{amsmath,amssymb,dsfont,epsfig,graphicx,xcolor,fontenc}
 \usepackage{ifpdf}
 \usepackage{amsmath}
 \usepackage{bm}
 \usepackage[english]{babel}
 \usepackage{graphicx}%
 \usepackage{amsfonts}%
 \usepackage{amssymb}
 \usepackage{braket}
\usepackage[normalem]{ulem}
 \usepackage{hyperref}
 \usepackage{epstopdf}
\definecolor{nicered}{rgb}{0.7,0.1,0.1}
\definecolor{nicegreen}{rgb}{0.1,0.5,0.1}
\definecolor{violet}{rgb}{0.7,0.3,0.3}
\hypersetup{colorlinks,citecolor= nicegreen,linkcolor= nicered}

\def\mysection#1{{{\vspace{2mm}\noindent\bf #1}~}}
\begin{document}

\def\LjubljanaFMF{Faculty of Mathematics and Physics, University of Ljubljana,
 Jadranska 19, 1000 Ljubljana, Slovenia }
\def\LjubljanaIJS{Jo\v zef Stefan Institute, Jamova 39, 1000 Ljubljana, Slovenia}
\def\Heidelberg{Institut f{\"u}r Theoretische Physik, Universit{\"a}t Heidelberg, Germany}

\title{Bump Hunting in Latent Space}

\author{Bla\v z Bortolato}
\email[Electronic address:]{blaz.bortolato@ijs.si} 
\affiliation{\LjubljanaIJS}
 
\author{Barry~M.~Dillon}
\email[Electronic address:]{dillon@thphys.uni-heidelberg.de} 
\affiliation{\Heidelberg}

\author{Jernej~F.~Kamenik}
\email[Electronic address:]{jernej.kamenik@cern.ch} 
\affiliation{\LjubljanaIJS}
\affiliation{\LjubljanaFMF}

\author{Aleks Smolkovi\v c}
\email[Electronic address:]{aleks.smolkovic@ijs.si} 
\affiliation{\LjubljanaIJS}

\begin{abstract}
\noindent Unsupervised anomaly-detection could be crucial in future analyses searching for rare phenomena in large datasets, as for example collected at the LHC.
To this end, we introduce a physics inspired variational autoencoder (VAE) architecture which performs competitively and robustly on the LHC Olympics Machine Learning Challenge datasets.
We demonstrate how embedding some physical observables directly into the VAE latent space, while at the same time keeping the anomaly-detection manifestly agnostic to them, can help to identify and characterise features in measured spectra as caused by the presence of anomalies in a dataset.
\end{abstract}

\date{\today} 

\maketitle

\section{Introduction}\label{sec:intro}
\noindent The absence of new physics (NP) discoveries thus far at the LHC strains many of the scenarios beyond the standard model (BSM) put forward in the last decades to address the theoretical and phenomenological weaknesses of the SM. 
It is possible that NP is present at mass scales just out of reach of the LHC, in which case effective field theory methods may help infer the presence of structures in the low-statistics tails of distributions measured at the LHC.
Another possibility however is that new degrees of freedom are already being produced at the LHC, but that existing search strategies have not been specific enough to disentangle their signatures from the backgrounds.
In going beyond the most motivated NP scenarios it becomes impractical to consider searching for all possible signatures.
To address this problem, unsupervised machine learning tools can be used to search for NP signals with no a-priori knowledge on what the relevant signatures may be. In particular, {\it anomaly-detection} techniques address the problem of searching for rare a-priori unknown signals in isolated regions of measured phase-space.
Several classes of unsupervised methods using Deep Neural Networks (DNNs) have been explored in the literature thus far; CWoLa-based methods 
\cite{Metodiev:2017vrx,Collins:2018epr,Collins:2019jip,1815227,Amram:2020ykb,Andreassen:2020nkr,Collins:2021nxn}\footnote{Most notably, the ATLAS collaboration has recently implemented a CWoLa-based weakly supervised di-jet search \cite{collaboration2020dijet}.}, 
autoencoder (AE) and variational AE (VAE) based methods 
\cite{Farina:2018fyg,Heimel:2018mkt,Roy:2019jae,Hajer:2018kqm,1800445,Alexander:2020mbx,Blance:2019ibf,Cerri:2018anq,Cheng:2020dal,Dillon:2021nxw,Finke:2021sdf,Atkinson:2021nlt}, 
and others
 \cite{Aguilar-Saavedra:2017rzt,Nachman:2020lpy,Mikuni:2020qds,vanBeekveld:2020txa,knapp2020adversarially,Khosa:2020qrz,Park:2020pak,Caron:2021wmq}.
However interpreting what has been learned by a DNN in physical terms is notoriously difficult. {Another class of unsupervised techniques based on Latent Dirichlet Allocation (LDA) \cite{Dillon:2019cqt,Dillon:2020quc,Blei03latentdirichlet} and other topic modelling methods \cite{Metodiev:2018ftz,Alvarez:2019knh} do provide interpretability, but are not based on DNNs.}
In this paper we introduce a novel optimization strategy and latent space sampling step to address two general outstanding issues in unsupervised VAE methods for (di-jet) anomaly-detection: (i) robustness of anomaly-detection performance, and (ii) physical characterisation of the VAE latent space.
Since we consider anomalies which are localised in the invariant mass of di-jets, the physical characterisation is in terms of both the observables used for anomaly-detection and the invariant mass of the events.

\section{Dataset and Observables}\label{sec:intro}
\noindent We demonstrate our approach using the LHC Olympics R\&D dataset, consisting of $10^6$ simulated QCD di-jet events and up to $10^5$ $Z'\rightarrow X(\rightarrow q\bar{q})Y(\rightarrow q\bar{q})$ events {(depending on the chosen signal to background (S/B) ratio benchmark)} with $m_{Z'}=3.5$ TeV, $m_{X}=500$ GeV and $m_{Y}=100$ GeV. 
Details on the simulation and kinematic cuts can be found in Ref.~\cite{LHCOrddata}. 
From each event we select the two jets with the highest $p_T$ and then order them by their mass in the input layer of the VAE.
For reasons of generality, we only use a small set of standard high-level observables in the analysis:\footnote{Observables more specially suited for the LHC Olympics data have been previously considered and can in principle result in further significant increases in performance for a specific dataset, albeit at the price of potential loss of generality, see e.g. Ref.~\cite{LHCObig}.}  the jet mass ($m_j$) and two ratios of N-subjettiness observables ($\tau_2/\tau_1,\tau_3/\tau_2$) for each jet~\cite{Thaler_2011}. 
For our chosen set of observables the signal (i.e. anomaly) events differ significantly from background in just the mass of the heaviest jet and $\tau_2/\tau_1$ of both jets, whereas the $\tau_3/\tau_2$ distributions exhibit significantly less distinguishing characteristics. 
This is realistic since in a typical model-agnostic BSM search we would expect only a subset of the observables to be sensitive to the signal. 

Using the VAE we want to obtain an anomaly score for each event, indicating how anomaly- or background-like the event is, which can then be used in a search for a localised excess (i.e. a {\it bump hunt}) in the di-jet invariant mass ($m_{jj}$) spectrum of the events.
Importantly in this approach, the di-jet invariant mass observable itself should not be among the inputs to the VAE, nor should it be directly computable from the inputs, since this {could sculpt the invariant mass distribution of the events passing the cuts leading to potential problems in quantifying the significance of any excess}
(see e.g. Ref.~\cite{Dolen:2016kst,Louppe:2016ylz,Shimmin:2017mfk,Moult:2017okx,Bradshaw:2019ipy,DiscoFever} for a more detailed discussion).

Finally, data pre-processing can have {significant} effects on the anomaly-detection performance, the stability, and the early stopping conditions.
We tested our method using different data pre-processing schemes available in the \texttt{scikit-learn v0.23.2} library~\cite{scikit-learn}.  We found that using the common \texttt{MaxAbsScaler} works well for our choice of observables, however we emphasise that this step should be handled with care, depending on the observables' distributions in a given dataset.

\section{Variational AutoEncoding}\label{sec:vae}

\begin{figure}[h!]
	\centering
	\includegraphics[width=0.5\linewidth]{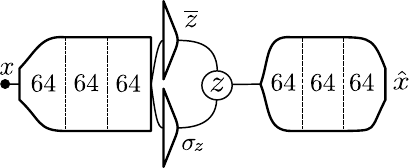}
	\caption{Standard encoder-decoder architecture used for the VAE, with $\bar{z}$ and $\sigma_z$ representing the mean and standard deviation that each event is mapped to by the encoder.
	\label{fig:NNScheme-1}}
\end{figure}

\noindent The VAE architecture~\cite{kingma2014autoencoding} and loss function define a probabilistic model for the di-jet data in which each event can be described by a single latent variable $z$.
The function $p(z| \text{event})$ is the posterior distribution and encodes information on the latent structure of that event.
As shown on Fig.~\ref{fig:NNScheme-1}, the VAE consists of two components, an encoder and a decoder, where the encoder models the posterior distribution for the model $p(z|\text{event})$ and the decoder models the likelihood $p(\text{event}|z)$.
The encoder consists of a neural network mapping each event to a mean $\bar{z}$ and a log variance $\log\sigma_z^2$, and a sampling step in which a value $z$ is sampled from a Gaussian distribution parameterized by $\bar{z}$ and $\log\sigma_z^2$.
The decoder consists of a neural network mapping the value $z$ back to a reconstructed event. 
Together this sequence defines a single forward pass through the VAE.
A key feature here is the bottleneck $z$, i.e. the latent space, which is a compressed representation of the event from which the decoder must attempt to reconstruct the full event. 
Note that for a severely compressed (low dimensional) latent space one does not expect the VAE to accurately reconstruct individual events, nor all the observable distributions. 
However, the medians of these distributions are expected to be learned most easily and thus reconstructed most accurately. 
The weights and biases of the VAE are optimised such that the Evidence-Lower-BOund (ELBO) of the probabilistic model evaluated on the dataset is maximised.
This ELBO consists of two terms, the reconstruction loss and the KL-divergence with respect to the normal prior in latent space, with the latter acting as a regulariser on the latent space distribution of events.
In general the two terms can be weighted differently relative to each other, which in the end relates to the variance on the reconstructed events produced on the output layer of the decoder~\cite{pol2020anomaly}.
The resulting loss function can be expressed as
\begin{equation}\label{loss}
\mathcal{L} = -\alpha\log p(\text{event}|z) - \tfrac{1}{2}\left( 1+ \log\sigma_z^2 - \bar{z}^2 - \sigma_z^2 \right)\,,
\end{equation}
where $-\log p(\text{event}|z)$ is modelled by the mean-squared-error between the input and output events, and the second term arises from the KL divergence between the posterior distribution for a single event and the Normal distribution.
Consistent with previous studies~\cite{Cerri:2018anq, pol2020anomaly, Cheng:2020dal} 
we find that $\alpha \gg 1$ results in good performance and avoids s.c. `component collapse' in which the KL divergence forces the means and log variances of all events to $0$.
For definiteness, in the following we present results for a value of $\alpha=5000$, however, our results were found not to be sensitive to changes in $\alpha$ in the range $\alpha \in [10^3, 10^4]$.

The weights and biases of the networks were trained via back-propagation with the following architecture: 3 hidden layers with 64 nodes and SeLU activations were used in both the encoder and the decoder, with the output layers of both having linear activations. 
Our final results are however insensitive to small changes in the choice of the DNN architecture or training parameters. 
All of the numerical procedures were implemented with \texttt{TensorFlow v2.3.1} \cite{abadi2016tensorflow}.\footnote{The code is publicly available at \url{https://github.com/alekssmolkovic/BuHuLaSpa}} 
In our analysis the networks were trained using a batch size of 1000 for up to a maximum of 100 epochs.

\section{anomaly-detection}\label{sec:results}

\noindent Traditionally the reconstruction loss has been used as an anomaly detection metric with autoencoders and even VAEs. 
In the latter however, there are other anomaly detection metrics that might be useful for detecting anomalous jets in the training data.
In particular the KL-divergence in the loss function between the (Gaussian) distribution of an encoded jet and the prior (unit Gaussian) distribution measures how much the encoded distribution of the jet deviates from the prior.
Since the KL term in the VAE loss function attempts to push all encoded distributions for the jets towards the prior, balanced against the reconstruction loss, this KL-divergence could be a good indicator of anomalous jets in the dataset.

\begin{figure}[h!]
	\centering
	\includegraphics[width=1.0\linewidth]{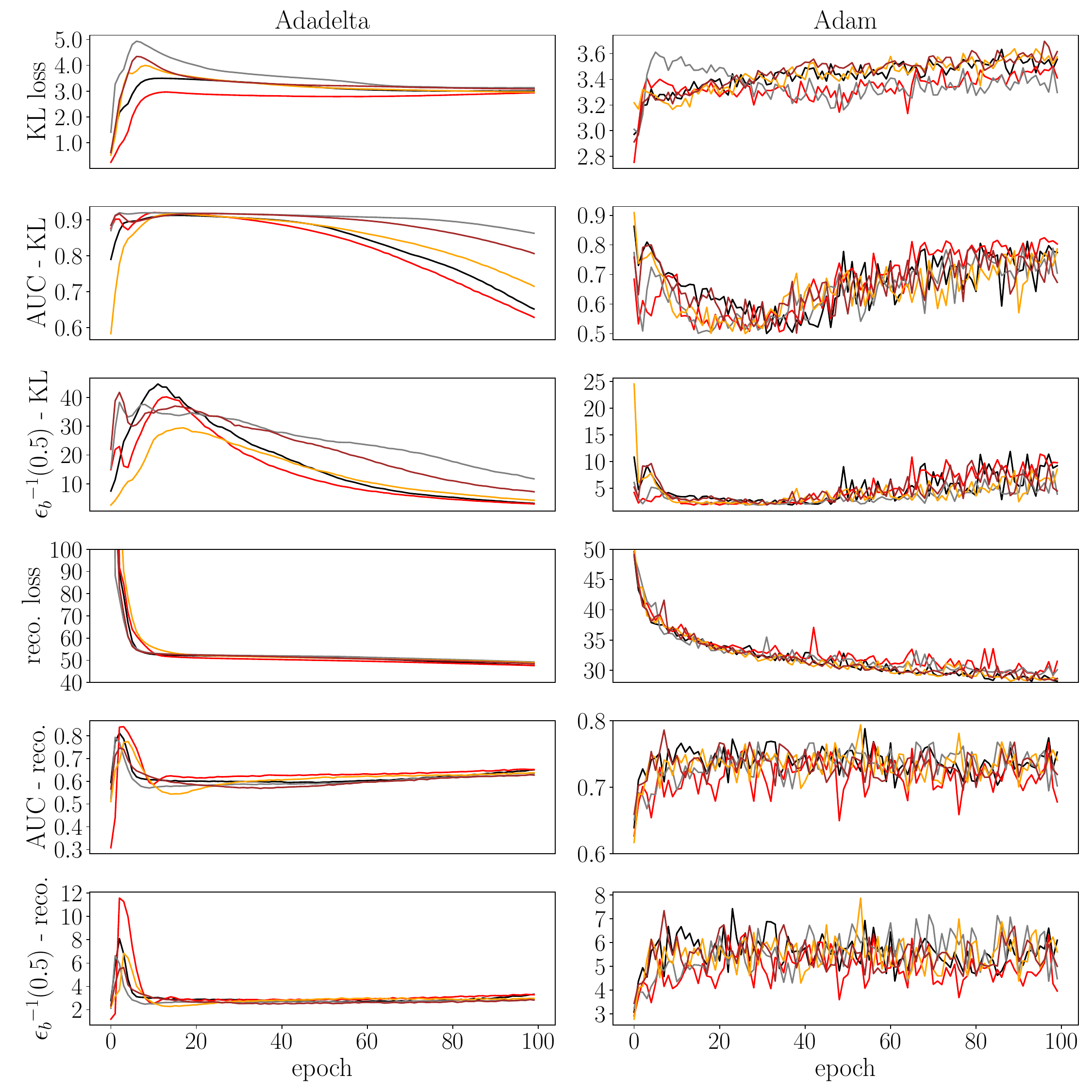}
	\caption{
{Comparison of various anomaly detection metrics as functions of the training epoch between 10 different trainings of the VAE, each using either the Adadelta (left column) or Adam (right column) optimiser. See text for details.}}
	\label{fig:adam-adadelta}
\end{figure}

In Fig.~\ref{fig:adam-adadelta} we present a detailed comparison of various anomaly detection metrics for 10 trained models, {differing only in the initial conditions of the weights,} as a function of the training epoch. The performance of different models is evaluated using the ROC area under curve (AOC) and the background mistag rate at signal efficiency of 0.5 ($\epsilon^{-1}_b(0.5)$).
For comparison we used the Adadelta optimiser \cite{zeiler2012adadelta} for half of the the models (left column), and the Adam optimiser \cite{zeiler2012adadelta} for the other half (right column), both with default hyper-parameters as implemented in \texttt{TensorFlow}.
{Note} the {unstable} performance of the results obtained using the Adam optimiser in comparison to those obtained with the Adadelta optimiser, {with both the loss function and the anomaly-detection performance fluctuating rapidly during training.}  {Adam includes the adaptive estimation of both first and second order moments when updating the network weights, and is designed specifically to increase the speed of training by reducing sensitivity to outliers in the data. Adadelta on the other hand only relies on second order moments, and uses adaptive learning rates per-dimension aiming to prevent the continuous decay of learning rates throughout the training while allowing the network to determine the learning rates on the fly.  In fact, by fine-tuning the hyperparameters of the Adam optimizer, we managed to reproduce the stability and performance that Adadelta provides. Nonetheless, in the following we use Adadelta, as the results are stable without having to resort to any fine-tuning. For an in-depth comparison of different optimisers, see e.g.~Ref.~\cite{DBLP:journals/corr/abs-1910-05446}}

{In the left column of Fig.~\ref{fig:adam-adadelta} we also see that the anomaly-detection performance using the KL-divergence as the anomaly detection metric is consistently better compared to using the reconstruction error for the same purpose. Moreover, we investigated this extensively and found that when the KL-divergence between the encoded validation events and the prior is at its largest, the performance of the KL-divergence as an anomaly detection metric on the testing data is at its best.}
We have only shown 5 example {runs} selected at random here, but the pattern persists in some form in all examples we checked.
As a cross-check we also studied the LHCO blackbox 1 data separately, with the results presented in \cite{LHCObig}, where we find the same behaviour as the study using the LHCO R\&D dataset here.
{From Fig.~\ref{fig:adam-adadelta} one can infer} on a possible reason for this behaviour being due to two terms (reconstruction loss and KL) in the loss function competing with each other in the minimisation procedure.
The KL term is there to regulate the latent space of the VAE, however it is possible that this term can over-regulate the latent space once the reconstruction-loss becomes small enough, leading to a decrease in the separability of the latent representations.
In light of these observations we propose a simple yet effective early-stopping procedure for the VAE.
We train the network until a peak is observed in the KL loss, and then terminate training and select the network at the epoch with the maximum KL loss.
We determine that a peak has been observed when the KL loss increases consistently for at least 5 epochs and then decreases consistently for at least 5 epochs.
With the selected network we then compute the latent representations of the events, and use the KL-divergence between the encoded event representations and the prior distribution as the anomaly detection metric.

\begin{figure}[h!]
	\centering
	\includegraphics[width=0.7\linewidth]{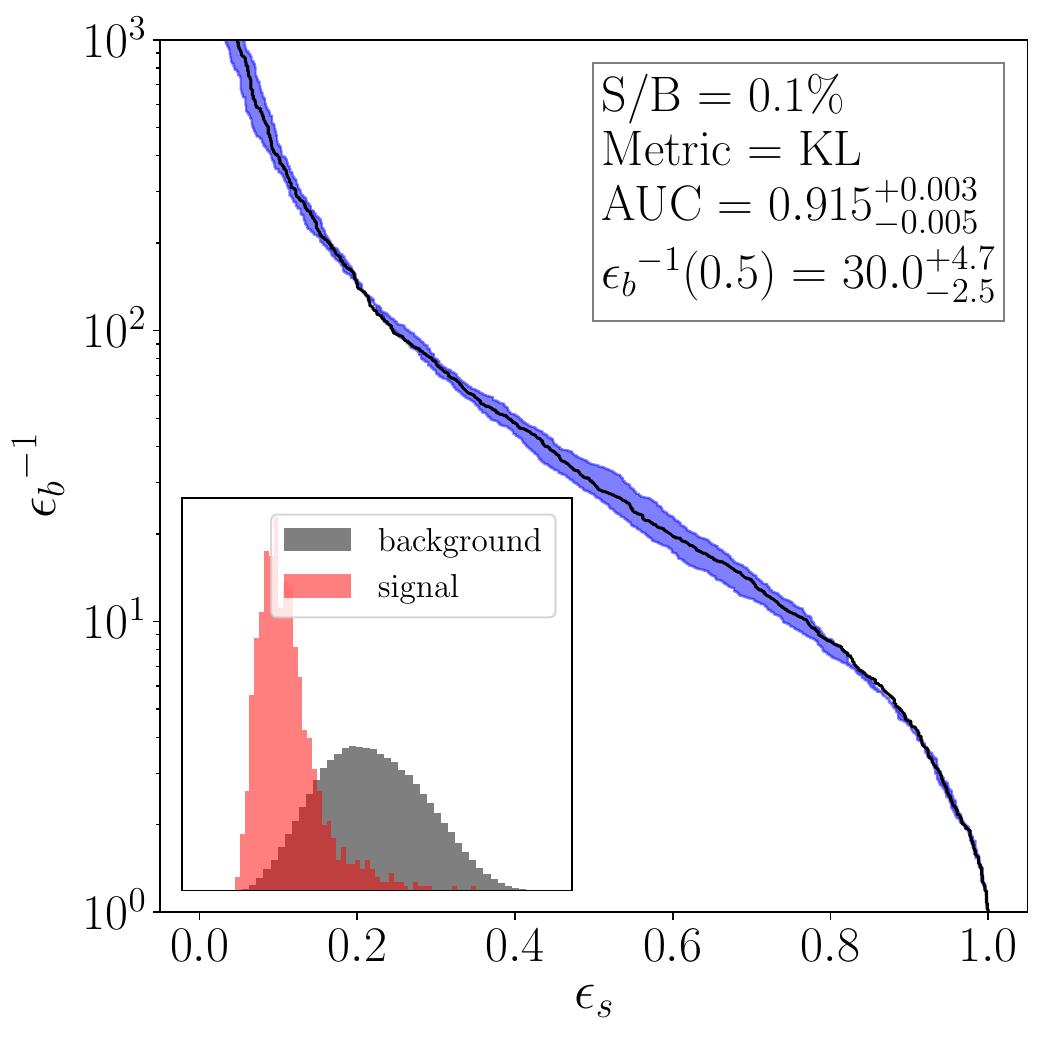}
	\caption{ROC performance on the LHC Olympics test data with $S/B=0.1\%$. {The uncertainty on the anomaly-detection, indicated by the blue region around the ROC curve, is estimated using the standard deviation of the per-event KL divergences around the mean over the 20 VAE training runs.  {The distribution of the KL metric for signal and background events is
shown in the inset plot.}} See text for details.}
	\label{fig:klclassifier}
\end{figure}

We considered 3 values of $S/B \in [10\%, 1\%, 0.1\%]$ and for each train 20 models {differing in initial conditions} according to the early-stopping prescription described above. 
For robustness we ensemble the output of the encoders in each of the 20 runs per $S/B$, using the mean of the per-event KL divergence as the anomaly detection metric. 
The performance of such an anomaly detection metric at $S/B=0.1\%$ is shown with a black line on Fig.~\ref{fig:klclassifier}, with the signal and background distributions of the average per-event KL shown in the inset plot. 
We do not show the results for larger $S/B$, however the only considerable difference is that the width of the blue band narrows for larger $S/B$, with the anomaly-detection performance remaining similar even at $10\%$. {In the anti-QCD tagger limit, i.e. training on a sample with background only events, we find that the classification performance is very similar to the case with $S/B=0.1\%$, albeit with increased uncertainty.}

\begin{figure}[h!]
	\centering
	\includegraphics[width=0.7\linewidth]{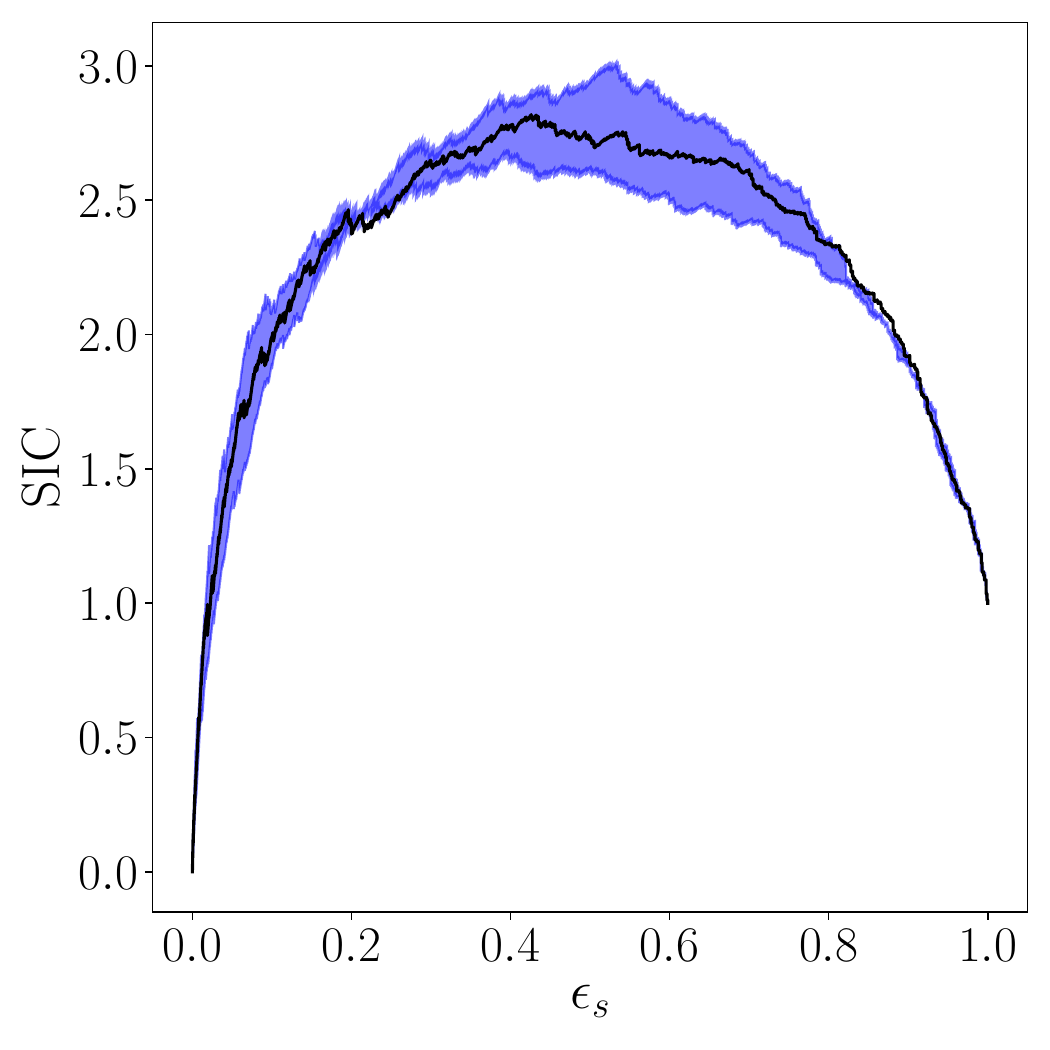}
	\caption{Significance-Improvement-Curve (SIC) for the VAE anomaly-detection on the LHC Olympics R\&D data with $S/B=0.1\%$.  The uncertainty on the curve is estimated {in the same way as in Fig.~\ref{fig:klclassifier}}.}
	\label{fig:sic}
\end{figure}

An important statistic in evaluating anomaly-detection methods is the improvement in significance ($S/\sqrt{B}$) before and after a cut on the anomaly detection metric.
We can see in Fig.~\ref{fig:sic} that with the VAE we achieve an improvement in the significance of {$2.75^{+0.21}_{-0.13}$} at a signal efficiency of $\sim\!0.5$.
This demonstrates clearly that the VAE is able to amplify the significance of the bump by separating signal and background events in latent space.
Note that this anomaly-detection technique does not employ sidebands nor signal region scanning to identify the anomalous events.
In practice the threshold on the anomaly score would be chosen such that a fixed number of events are allowed to pass the cut \cite{collaboration2020dijet}.
For a comparison of these results to other methods tested on the same dataset, we refer the reader to Ref.~\cite{LHCObig}.
\section{Latent Space Characterisation}\label{sec:latchar}

\noindent Understanding what information deep-learning tools use to arrive at a particular outcome is generally of great interest in physics applications.
In the case of anomaly-detection, we would like to be able to identify why a particular event was assigned a high anomaly score.
The simplest way to do this is to inspect the observables for events with large anomaly scores, or in this case we could also inspect events in different regions of latent space, which is especially simple when the dimension of the latent space is small.
In this section however we propose a different strategy.

In many anomaly-detection cases, including the example studied here, the anomalous events will form a resonance in the invariant mass variable.
These types of signals can be efficiently searched for in a bump hunt analysis.
In a bump hunt, {signal regions and sideband regions} are defined by cuts in the invariant mass, with the sidebands used to estimate the amount of background in the signal region.
Comparing with the measured number of events in the signal region a limit on a potential signal cross-section can be estimated.
Using the anomaly scores calculated from the VAE we could pre-select events for the bump hunt by imposing a cut on the KL-divergence for each event.
This could significantly increase the significance of the signal events in the sample, while at the same time significantly reducing the total number of events in the analysis.
In order to ensure that this pre-selection does not artificially sculpt some feature in the invariant mass distribution we select observables for the training of the VAE that do not contain information on the invariant mass of the events.
Most importantly, we do not include the invariant mass observable in the input to the encoder.\footnote{{In practice it is unfeasible to select observables with absolutely no indirect correlation with the invariant mass. Known decorrelation techniques can be employed to reduce residual correlation effects, see e.g. Refs.~\cite{Dolen:2016kst,Shimmin:2017mfk,Moult:2017okx,Bradshaw:2019ipy}}.}
The latent parameters for each event can be written $\bar{z}_i(\{\mathcal{O}\})$ and $\sigma_{z,i}(\{\mathcal{O}\})$, with $\{\mathcal{O}\}$ being the jet mass and N-subjettiness observables used in training the VAE.
Once we have acquired the anomaly scores, we want to gain some physical insight into what is, and what is not, encoded in the latent space.

\begin{figure}[h!]
	\centering
	\includegraphics[width=0.5\linewidth]{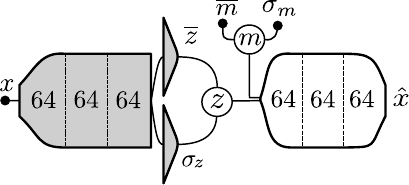}
	\caption{Modified sampling and decoder architecture for the characterisation of the latent space as a function of the invariant mass of the di-jet events.
	Since only the decoder is (re)trained at this stage, the greyed out region indicates that the weights of the encoder are frozen.
	\label{fig:NNScheme-2}}
\end{figure}

We do this by training a separate decoder network using the means and variances from the encoder network trained to provide the anomaly scores.
Thus the weights and biases in the encoder network are frozen for this portion of the analysis.
The key difference is that the new decoder is trained with the invariant mass information from each event.
The inputs to the decoder for each event are $z\!\sim\! \mathcal{N}\left(\bar{z}_i,\sigma_{z,i}\right)$ and $m\!\sim\!\mathcal{N}\left(m_{jj},\sigma_m\right)$, with $m_{jj}$ being the invariant mass and $\sigma_m$  an estimate of the uncertainty on the invariant mass (for concreteness we fix its value to $\sigma_m\!=\!0.025m_{jj}$).
This can be thought of as extending the latent space of the trained encoder to now include an invariant-mass direction. We also use a \texttt{StandardScaler} pre-processing step on the invariant masses.
The architecture of this separate decoder is shown in Fig.~\ref{fig:NNScheme-2}.
The loss function now only contains a reconstruction term, since the latent space parameters are fixed.
So the decoder is again optimised to minimise the reconstruction of the events in the training data, but now with the additional invariant mass information.
If this new information is useful to the decoder in improving the reconstruction of the signal events, we should be able to see imprints of the signal features when plotting a 2D histogram of the {reconstructed} observables $\mathcal{O}$ with the $(x,y)$ axis being $(m_{jj},\bar{z})$.
These imprints could provide hints to the invariant mass of any anomalous clusters of events in the dataset, and also indicate where in the space of $\{\mathcal{O}\}$ these events are.

\begin{figure}[h!]
	\centering
	\includegraphics[width=0.7\linewidth]{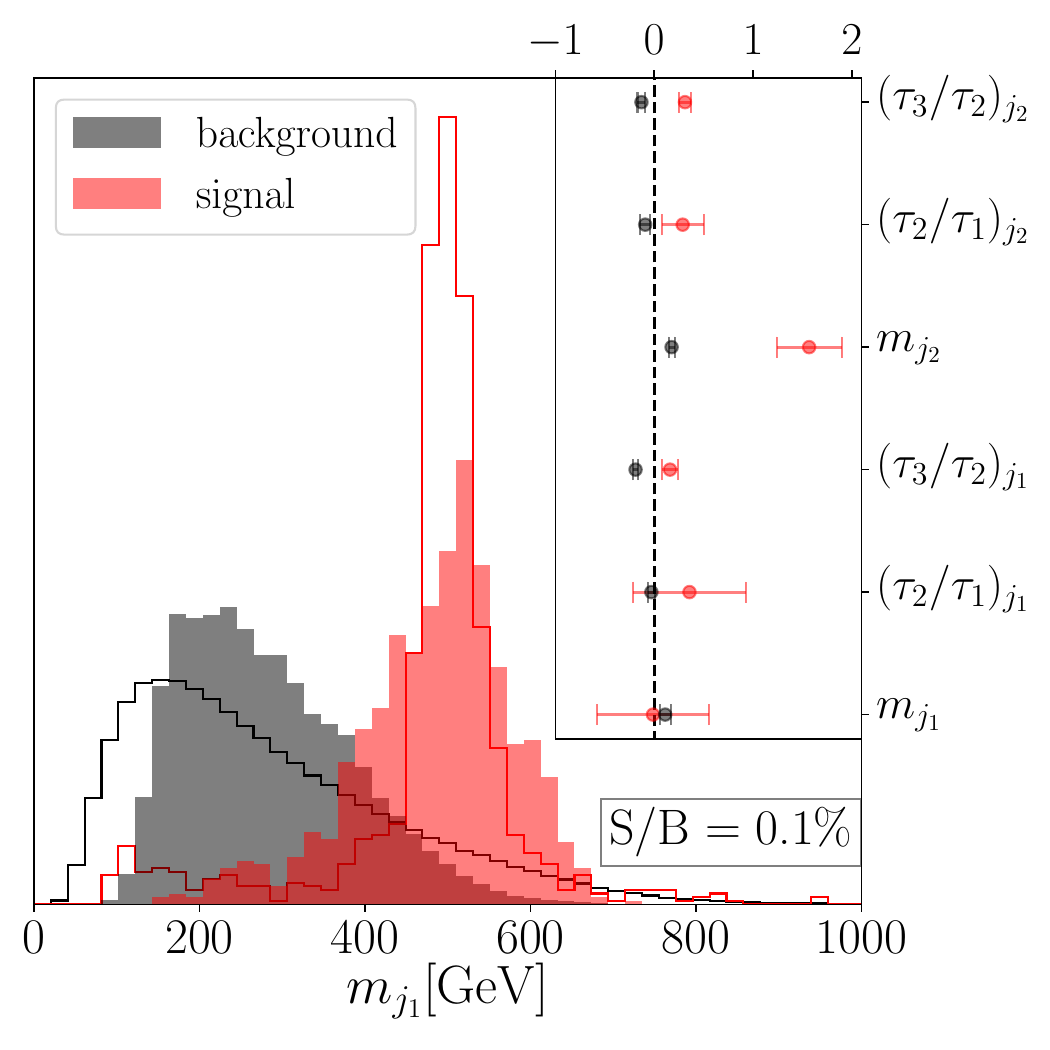}
	\caption{Input (outline) and reconstructed (filled) leading jet mass distribution for both signal and background, with the inset plot showing the difference between medians of the input and reconstructed distributions, normalized to the standard deviations of the input distributions. {The error bars indicate a sigma variation in the predicted medians averaged over the 20 VAE training runs.} See text for details.}
	\label{fig:1D_reconstruction}
\end{figure}

In training the separate decoder to interpret the latent space, we find that the Adam optimiser provides better results than the Adadelta optimiser used in the anomaly-detection step.
This is because the latent representations are fixed and we now only have the reconstruction loss to optimise, so we focus on obtaining as best reconstructions of the jet observables as possible. We therefore train the decoder until the reconstruction loss converges.
In Fig.~\ref{fig:1D_reconstruction} we plot the reconstruction of the leading jet mass for both signal and background, while in the inset figure we plot the difference between the median of the reconstructed distribution and the median of the input distribution for each observable, normalized by the standard deviation of the input distribution.
Despite having just $S/B=0.1\%$ the network is able to reconstruct some of the main distinguishing features of the leading jet mass distribution for both signal and background, and satisfactorily reconstructs all other observables' medians with the exception of the lighter jet mass, which is however also less than $2\sigma$ away from its input value.
This is to be expected, since we use just a 1D latent space, the amount of information on the event that can be encoded here is very limited.

\begin{figure}[h!]
	\centering
	\includegraphics[width=1\linewidth]{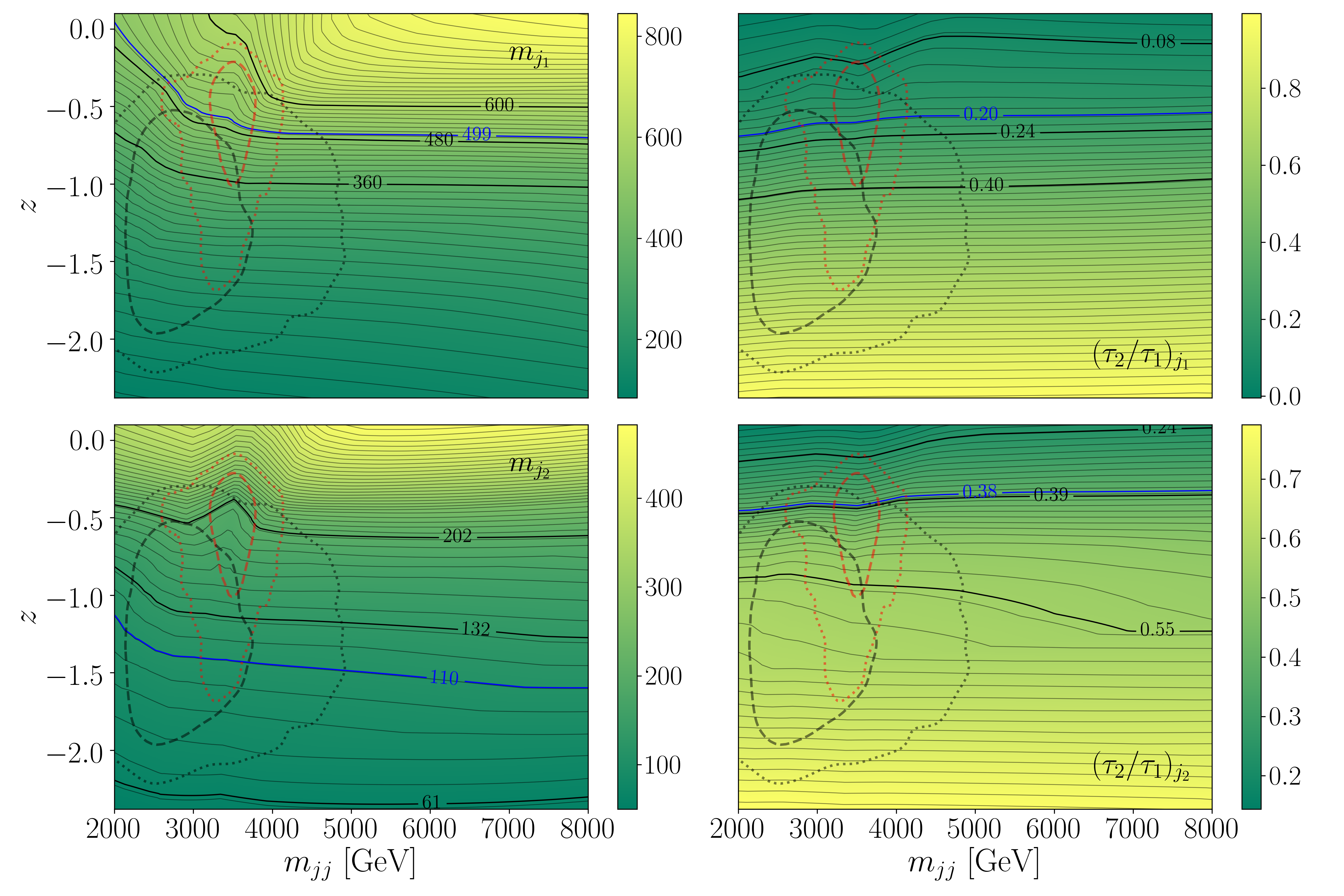}
	\caption{Generated jet observables ($m_j$, $\tau_2/\tau_1$) produced by scanning over the the latent space observables $(z,m_{jj})$ in a network trained with $S/B=1\%$.  For each feature the solid black contours denote the values of the true signal mean and one standard deviation around it, while the solid blue contour shows the value of the median. The dashed and dotted contours show the regions in the latent space in which $68\%$ and $95\%$ of signal (red) and background (black) resides.}
	\label{fig:heatmap}
\end{figure}

Since we can treat the latent space as a function of just two variables, the values $z$ and $m_{jj}$, we can visualise the reconstructed observables in a series of 2D heatmap plots.
We select the ranges $z\in [-2.5,0]$ and $m_{jj}\in [2,8]~\mathrm{TeV}$ based on where the events are mapped to in latent space and their invariant masses.
We then construct a grid on this 2D space, passing each point through the newly trained decoder to obtain a reconstructed event.
To demonstrate this we focus on a scenario with $S/B=1\%$.
The reconstructed observables for the events are plotted as 2D heatmaps in Fig.~\ref{fig:heatmap}.
Note again that the $z$ direction in these plots is approximately independent of the invariant mass, due to the selection of observables used to train the autoencoder in the first step.
To aid the interpretation of these plots we have added contour lines so that the differences in the relative magnitude of the observables are clearer.
We have also overlaid contours to indicate where in the latent space the background events and anomalous events are, so that correlations with the reconstructed observables are clearer.
With this, we can see that there is indeed a strong correlation between the localized {features in the} invariant mass {contours} in latent space at $\sim3.5$\,TeV, and the actual location of the anomalous events in latent space.
This a first indication that the invariant mass sampling step does provide additional information to aid the reconstruction of the anomalous events.

\begin{figure}[h!]
	\centering
	\includegraphics[width=1\linewidth]{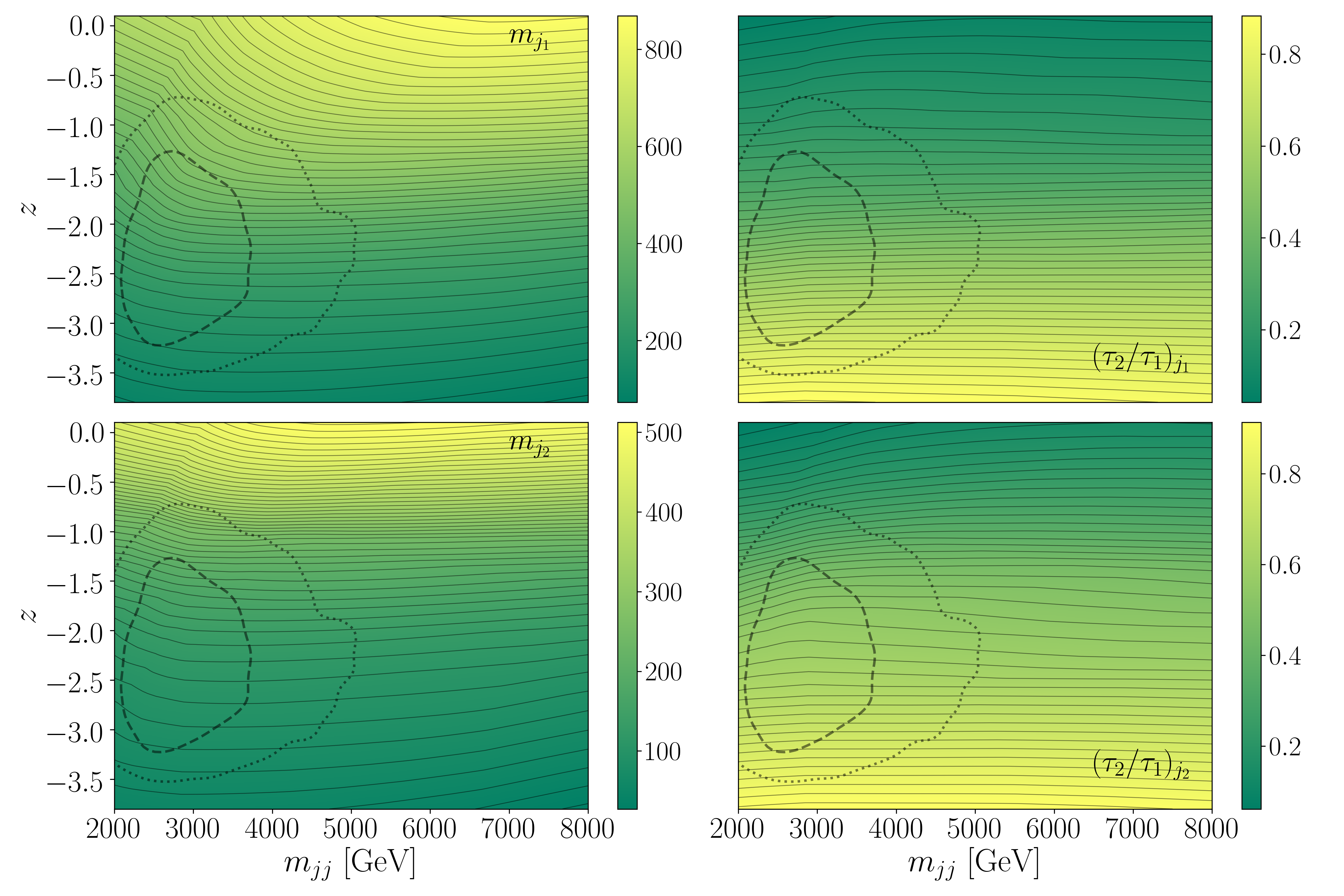}
	\caption{Generated jet observables ($m_j$, $\tau_2/\tau_1$) produced by scanning over the the latent space observables $(z,m_{jj})$ in a network trained with no anomalies, $S/B=0\%$.  {See caption of Fig.~\ref{fig:heatmap}} for details.}
	\label{fig:heatmap-nosignal}
\end{figure}

In order to evaluate the significance of these results, we have run {the same two-step} training procedure but with no anomalies in the dataset, $S/B=0\%$, and plotted the results in Fig.~\ref{fig:heatmap-nosignal}. Here we can clearly see that there are no significant localized features in the reconstructed observables over the $z-m_{jj}$ plane {where $m_{jj}\!\gtrsim\!3$ TeV.}
In both Fig.~\ref{fig:heatmap} and Fig.~\ref{fig:heatmap-nosignal} there is a slight correlation between $z$ and $m_{jj}$ {for $m_{jj}\!\lesssim\!3$ TeV}, that is strongest in $m_{j_1}$.
This is due to a residual correlation between $m_{j_1}$ and the di-jet invariant mass present in any finite jet $p_T$ range. 
However, since the signal induced localized features {in the invariant mass contours} in latent space are clearly independent of these correlations, our results could potentially be further improved through the use of techniques to remove the correlations between $z$ and $m_{jj}$ (for the background).

\begin{figure}[h!]
	\centering
	\includegraphics[width=1\linewidth]{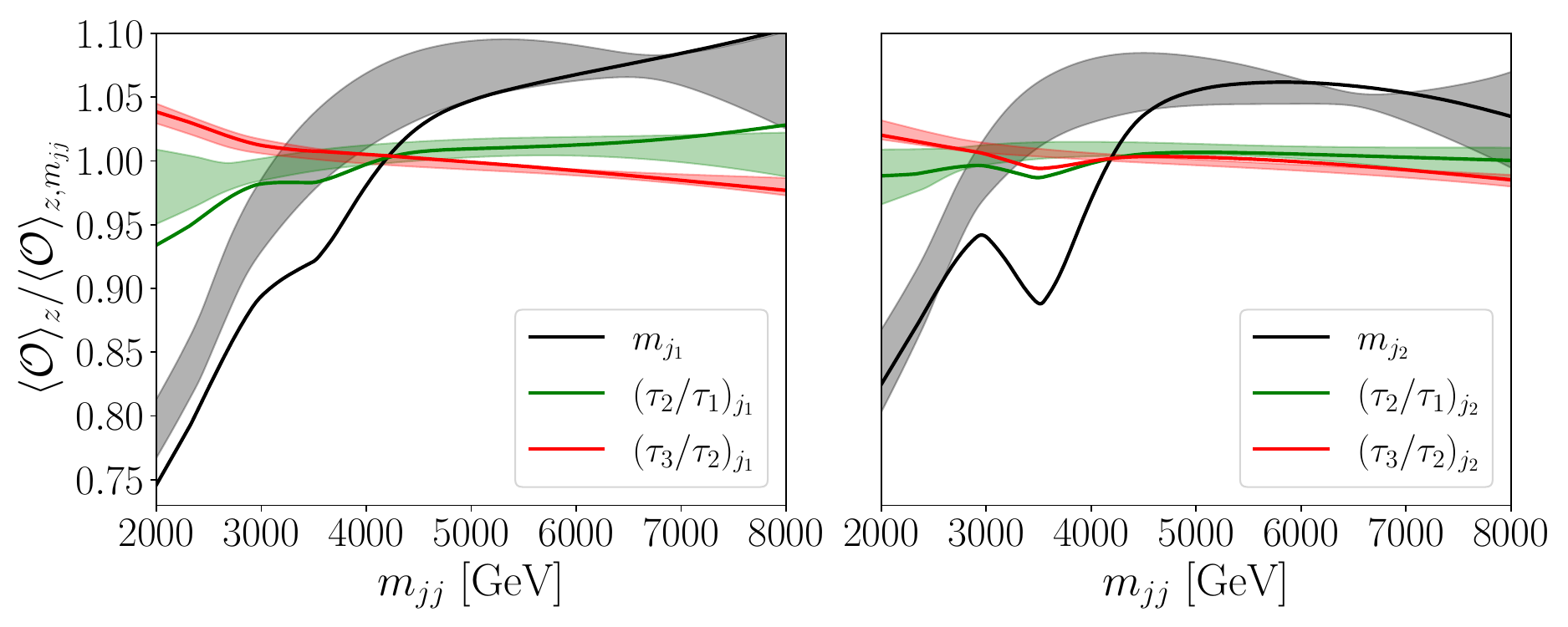}
	\caption{Marginalized distributions of observables ($\langle \mathcal O\rangle_z / \langle \mathcal O\rangle_{z,m_{jj}}$) computed for random subsets of background-only $S/B=0\%$ events (shaded bands) and a dataset with $S/B=1\%$ (full lines). See text for details.}
	\label{fig:marginal}
\end{figure}

In addition it is important to be able to distinguish  imprints of actual anomalies localised in invariant mass from possible statistical fluctuations in the $z-m_{jj}$ distributions.
We quantify effects of statistical fluctuations in the data by training on 10 random subsets of background-only events with 900k events each.
We then compute the marginalized distributions of observables, $\langle \mathcal O\rangle_z$, where $\mathcal O$ is averaged over $z$ for a fixed value of $m_{jj}$, normalized to the average value of $\mathcal O$ over the whole $(z, m_{jj})$ range $( \langle \mathcal O\rangle_{z,m_{jj}})$. The results  are shown in Fig.~\ref{fig:marginal}, where the range of results of the subsampled background runs are shown as shaded regions. On the other hand the full lines depict results in the presence of the signal with $S/B=1\%$. 
Note that deviations from $\langle \mathcal O\rangle_{z} / \langle \mathcal O\rangle_{z,m_{jj}} \simeq 1$ at low $m_{jj}$ are again due to aforementioned slight correlations between $z$ and $m_{jj}$.
However, more importantly, we observe significant localized deviations from the smooth background-only bands in the signal region of $m_{jj}$, making it clear that localized deformations in the contours imply the presence of anomalous events, distinguishable from random statistical fluctuations in the data (or residual $z - m_{jj}$ correlations). 

Lastly we would like to comment on the current limitation of this interpretation technique. Here we have shown results for $S/B=1\%$. 
For smaller numbers of anomalous events the signal features {in the invariant mass contours} in the latent space become less prominent and thus more difficult to disentangle from possible fluctuations or residual (encoder induced) $z$ -- $m_{jj}$ correlations. In the future we plan to explore ways to mitigate this and extend the application of our technique to work in more realistic S/B scenarios.

\section{Outlook}\label{sec:outlook}
\noindent In this paper we have introduced an effective approach to anomaly-detection using information encoded in the latent space of a VAE.
We then discussed a novel method for the characterisation of  anomalies in a given dataset which can aid in the interpretation of results from a bump hunt analysis.
Our method introduces `bumps' {(in the invariant mass contours)} in latent space, which aid in the interpretation of  anomalous events as localised bumps in the (invariant mass) spectrum and could aid in defining signal windows in a prospective experimental analysis.
The performance of the anomaly-detection and the effectiveness of the characterisation are demonstrated on the LHC Olympics R\&D dataset. 
We also applied it successfully to the Black Box 1 dataset, where it compares favourably to other existing anomaly-detection approaches, see Ref.~\cite{LHCObig}, thus demonstrating its robustness. 
Our method {(apart from the choice of observables - i.e. the feature space)} is however general and could be applied to other physics datasets beyond LHC di-jet spectra. 
The approach could potentially be refined in several directions, either by  (1)  enlarging the number of latent space dimensions encoding observables used for anomaly-detection, particularly with the goal of increasing the sensitivity of the physical latent space characterisation to smaller S/B ratios, as well as improving the observables' reconstruction accuracy; by (2) incorporating additional (uncorrelated) scanning observables used in the characterisation step (in addition to $m_{jj}$), with the aim of incorporating physical features into the latent space in order to better separate signal and background encodings; finally (3) the characterisation step could potentially be used to help define signal windows in more realistic NP search analyses, all of which we leave for future work.

\mysection{Acknowledgements} The authors thank Andrej Matevc for his involvement in the initial stages of the project. BB, JFK and AS acknowledge the financial support from the Slovenian Research Agency
(grant No. J1-3013 and research core funding No. P1-0035). BMD acknowledges funding from BMBF. 

\bibliographystyle{JHEP}
\bibliography{current}

\end{document}